\documentclass[prb,preprint,superscriptaddress,showpacs]{revtex4}
\usepackage{revsymb}
\usepackage[dvips]{graphicx}
\usepackage{color}
\usepackage{here}
\usepackage{amsfonts}

\begin{document}

\title{ {\Large Social networks from communities of electronic mail} } 

\author{ {\large {\bf Horacio Castellini}} }
\email{horacio9573@yahoo.com.ar}
\affiliation{{\rm Dpto\@. de F\'{\i}sica, F\@.C\@.E\@.I\@.A\@., 
Pellegini 250,
2000 Rosario}}

\author{{\large {\bf Lilia Romanelli}}}
\email{lili@ungs.edu.ar} 
\affiliation{{\rm Instituto de 
Ciencias, Universidad de General
Sarmiento, J.M.Gutierrez 1150, Los Polvorines (1611),Buenos Aires 
Argentina}}

\begin{abstract}

Social networks are analyzed as graphs under the scope of discrete mathematics which 
have a great range of applications in different
contexts such as: technology, social phenomena and biological systems. 
At the present this theory gives a set of tools for a phenomenological analysis that 
would be difficult or almost impossible with a different 
approach. In this work social networks for different technical communities from 
electronic mail and ``News'' in Spanish language are constructed. The algorithm was
based on the use of RFC2822 standards and RFC1036 to arm threads of messages.
The results are quite different from that obtained by  another kind of community as 
the jazz musicians community. Nevertheless they show an analogy to random graphs 
obtained by the ``Configuration Model'' method.
This points the attention that some generalization assumptions are not correct.

\end{abstract}

\maketitle

\section{Introduction}

Complex social networks associated to {\em Internet}, like as e-mail lists, 
news services etc\@. do not have a structured architecture like a project in network 
communication engineering. 
They show some kind of synergia given by the great amount of users who form the 
mentioned community. This study faces some problems, some time neither cultural 
guidelines are taken into account and the results are generalized with quite 
different guidelines.
This is why we have restricted our analysis to one language and to the technical 
communities as an experimental application of the theoretical tools in social networks.
This work is organized in two sections: the first one is devoted to a brief 
introduction on social networks and the other is referred to the study of 
communities coming from e-mail and ``News Services'' in Spanish language.

\subsection{Networks}
A social network is a set of relations (links or edges) among different elements 
(nodes, vertices or actors). Formally a network is a graph
$G=(V,E,\gamma)$ where $V$ is the set of vertices, $E$ is the set of edges 
and $\gamma:E \to V$ such $\gamma(e)=\{v,w\}$. 
That means, $\gamma(.)$ to each edge a pair 
of vertices are assigned and they are known as ends of edge.
Recently for the the networks study binary matrices are used, therefore an isomorphism 
exists $f:G \to B_n$, where $B_n$ is the set of square binary matrices of 
dimension {\it nxn}. This matrix is called adjacency matrix (AM)\cite{r1}.
Social scientists defined by convention that actors (output, egos) are placed in the 
rows, while the attributes or related actors (input, alter) in the columns.
The AM is symmetric since we are dealing with non directed graphs\cite{r2}. 
There are {\it multiple} graphs in which more than one kind of edges are 
identified as: (kinship, friendship); (theme, author); etc\@. 
this would be quite useful for building social substructures although for building 
the required AM is more complicated. Usually we regard an associated AM with some 
particular projection i\@.e\@. {\em kinship}.
Otherwise graphs may be weighted, that means it is possible to assign a weight to each 
link. This gives us a non zero value associated to each AM element.
As can be seen the AM has all sensitive information related with the social network 
in particular. It is worth to notice that the diagonal elements in the matrix are 
filled with zeros or are neglected in the algorithms since the self interaction has 
no sense. However, then have to be considered in their booleans products \cite{r3}.
On the other hand some other properties can be associated with the AM.
``Macroscopic'' properties among the actors as {\bf Path}: Is a concatenation 
of vertices connected by edges so that no chosen edge is repeated  
$\pi=\{ x_1, x_2, \cdots, x_n \}$ where $x_1$ is the initial vertex and 
$x_n$ the final vertex\footnote{Where the length is the number of vertices}.
{\bf Geodesic}: is the path of minimal length\footnote{May be not unique} 
among actors. 
If it is not exist, as in the case of the non connected graphs, the infinite value is 
taken like the length.  The geodesic path is the optimal path between actors, because 
socially, are actors with strongest links.
Also there are ``Microscopic'' properties as {\bf clique}. 
This is a measurement of the transitive triades of the network.  
Two definitions exist, an originating one of theory of graphs. 
Which knows it like transitivity a graph:
\begin{displaymath}
T(G)=\frac{3 \, \times \, \textrm{\# de triangles}}
{\textrm{\# de triplets}}
\end{displaymath}
where $\#$ is the cardinality of the set and the other one was formulated by 
Watts y Strogatz, known as {\bf clustering} associated with the actor $i$ 
is defined as:
\begin{displaymath}
C_i(G)=\frac{\textrm{ \# of edges in i}}{\textrm{Maxima edges\# in i}}
\end{displaymath}
and then averaged over the actors $C(G)=<C_i>$ this is the ``clustering''\cite{r4}.

As much in a case as in another one, the transitive triades are small groups that 
represent the balance or the natural state towards which tend the social relations.
But in either case they are small groups.
Another ``Microscopic'' property is the average degree of connection among the closest 
neighbors (CN)\cite{r5} defined as:

\begin{equation}
K_{nn}(k)=\sum_{k'} k' \, P(k/k')
\end{equation}

where $P(k'/k)$ is the conditional probability the a vertex of $k$ degree is connected 
with another one with $k'$ degree. On the other hand when $K_{nn}(k)$ is an increasing 
function the network is called {\em associative}, and if is decreasing is called 
{\em dissociative}. Also we may characterize the network from rows histograms, 
known as {\bf prestige of the actor} this is coincidently with the columns 
histogram known {\bf popularity of the actor}. 
But at the moment another kind of statistics is used 
called the {\em connectivity} probability, $P(k)$. $P(k)$ is the probability that a 
randomly chosen vertex have $k$ edges \cite{r6}.

According with the functional shape of the histogram's tail
($k \to \infty$) the network can be classified as 
{\bf exponentials}, when $P(k) \sim e^{-\lambda k}$; 
{\bf scale-free}, when $P(k) \sim k^{-2-\gamma}$ with $\gamma > 0$;
{\bf broad-scale}, when is ``scale-free'' but with an abrupt cutoff; and 
{\bf single-scale}, when has a fast asymptotic decay \cite{r7}.
In mostly field experiments the {\bf scale-free} network was shown as the dominant. 
In order to estimate the exponent the cumulative probability $P(k'>k)$ is 
used\cite{r8} defined as:

\begin{equation}
F(k)=P(k'>k)=\sum_{k'>k} P(k')
\end{equation}
if $P(k) \sim k^{-\gamma}$ in the tail, then 
$F(k) \sim \int P(k')\,dk \sim k^{-\gamma+1}$.

\section{Application to e-mail Spanish lists}

All the non ponderated properties of the relation among actors are included in the AM. 
Each index row is associated with the actor who generates the subject, 
{\bf author root} and each column index with the actors involved in the thread of 
conversation {\bf author descendent}. A lexicographical arrangement is not used but by 
prestige\cite{r9}, in other words, a low index is assigned to the actor with the 
greatest absolute frequency and successively until the index of greater value 
corresponding to the author of less prestige.
For the construction of the matrix we have not taken into account the self-answers 
and no the threaded demands (without thread). Due to this fact they are discarded in 
previous phases to the application of the algorithm.

\subsection{Algorithm used for the message threading}
According with the standard RFC 822 and derivate RFC 2822 and 1036\cite{r10}
the transmission format of the messages coming from  electronic mail and News 
services should be composed by some headers fields and a body in plain text.
From all the fields sent in a single message the following fields are required to 
construct threads of messages::

\begin{description}
\item[From:] This field contains the identity and direction of the person who 
sends the message
\item[Subject:] This field indicates the nature of the message.
\item[Message-ID:] This field must be unique for each message
the suggested format is ``$<$local\_part@domain$>$''.
\item[In-Reply-To:] refers to a or the {\bf Menssaje-ID} 
where is the message is answered, if the message is new this field does not appear.
\end{description}

Before the use of threading algorithm, the messages go through a filter 
which extract the field of interest and delete unwanted message.
This filter is written in{\em PERL}\cite{r11}. 
Since mostly of the codes for messages threading are not 
free domain it was necessary to write our own code in ``C'' 
language from an existing one GPL\cite{r12} in ``JAVA''language, 
modified for generating a list from the actors sequence related with a thread, 
having as the list root the actor whose give the 
beginning of the thread. This algorithm have prove its robustness in hundred 
thousand trials. 

A brief sketch of the algorithm is given.
The algorithm is based on the handling of connected structures of data which are:

\begin{verbatim}
Container{
     Message message;
     Container parent;
     Container child;
     Container next;
};
\end{verbatim}

The field ``message'', may be {\bf NIL}. The structure ``Message'' 
have the following fields:

\begin{verbatim}
Mensaje{
     char* Subject;
     char* Message_ID;
     char* In-Reply-To;
     char* From;
};
\end{verbatim}

When the field  ``In-Reply-To'', Take the value {\bf NIL} is indicating the 
message father.
An indexed table is associated where in index is ``Message\_ID'' from the message 
parent.
Then a ``Container'' root or parent is associated. After that using the threading 
algorithm a message data base descendent associated is built. 
A table ordered by absolute frequencies of appearance of the "Author" of the 
message father is generated with decreasing order.
Finally a AM is built from the previous results. For algorithm details see
``{\em Message Threading of Jamie Zawinsky, technical report}''\footnote{On Internet}.

\subsection{Analysis of the obtained Adjacency Matrices}
\begin{figure}[h]                                                               
\begin{center}                                                                  
\includegraphics[width=12cm, height=13cm,angle=-90]{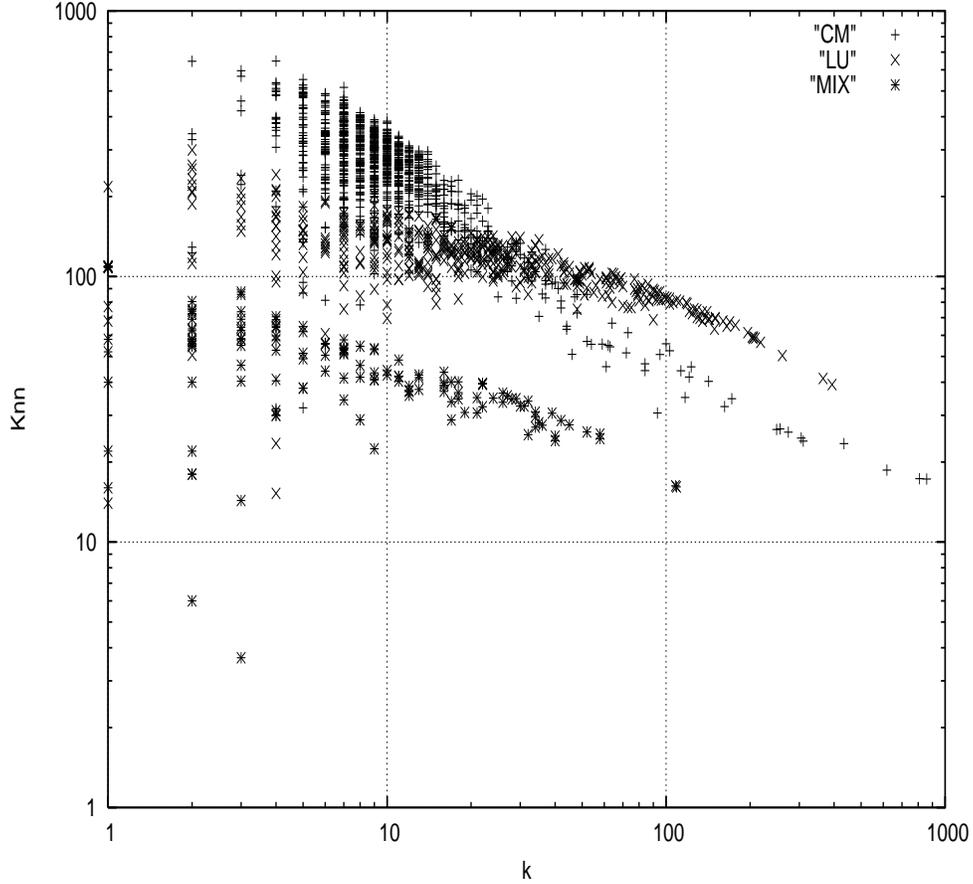}
\end{center}                                                                    
\caption{Plot of $K_{nn}(k)$, {\bf CM} coming from the  
``{\em Configuration Model }'' algorithm, {\bf LU} corresponds to purely technical list
and {\bf MIX} to a list of general interest.
The data were not averaged  intentionally in order to get a better picture 
from the scattering of them.}
\label{fig:1}                                    
\end{figure}

We have taken as leading cases two mailing lists in Spanish which represents 
the observations done in previous studies. 
One is a purely technical list called {\bf LU} 
and another one with the same actors but with a more general thematic called {\bf MIX}.
This work have as null hypothesis those obtained from the algorithm 
``{\em Configuration Model }'' (CM)\cite{r5} composed by 1200 vertices.
This allow us to build random networks with a given probability density of edges 
by vertex, $P(k)$. We have used $P(k)\sim k^{-\gamma}$ with $\gamma=2.1$, since for 
this value stands the behavior for $k\to\infty$ for the null hypothesis and the 
real cases.
As can be observed in fig-\ref{fig:1} in both cases the asymptotic behavior of 
$k_{nn}$ is agree with those proposed by the CM. This dissociated mixed behavior is 
quite different to those found in jazz communities\cite{r13} or scientific 
collaborators network\cite{r6}, due their behavior is purely mixed associative.

The following is a comparative table between the calculated values of ``{\em clique}'',
 {\bf C} by using  Watts and  Strogatz's algorithm and the averaged geodesic 
{\bf G} for each case.

\begin{center}
\begin{tabular}{|l|c|c|c|}
\hline
 & {\bf CM} & {\bf LU} & {\bf MIX} \\
\hline
{\bf C} & $0.50$ & $0.7 \pm 0.3$ & $0.9 \pm 0.2 $ \\
\hline
{\bf G} & $3.63$ & $ 3.36 \pm 0.02$ & $2.83 \pm 0.02$\\
\hline
\end{tabular}
\end{center}

\begin{figure}[h]                                                               
\begin{center}                                                                  
\includegraphics[width=12cm, height=13cm,angle=-90]{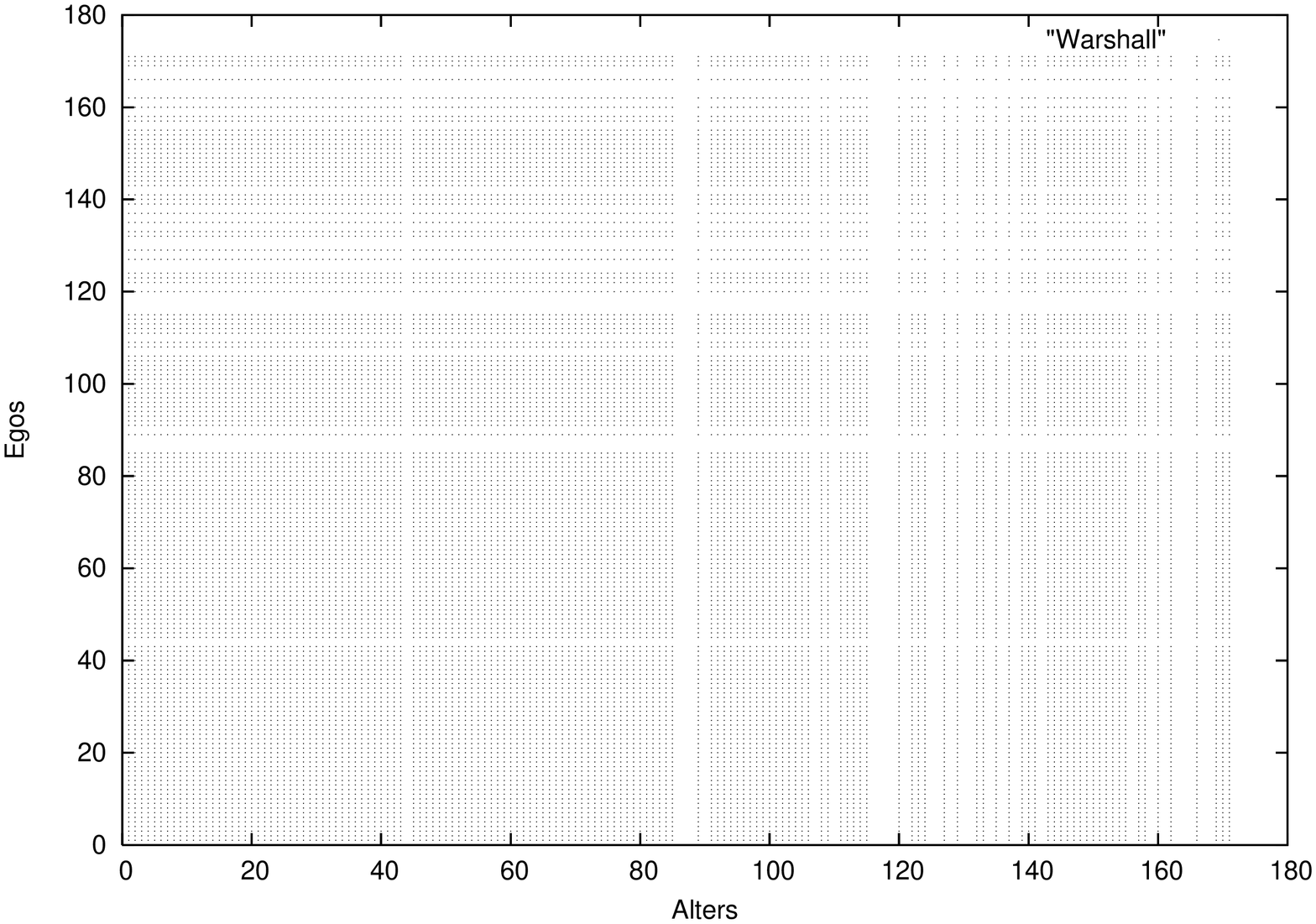}
\end{center}                                                                    
\caption{Plot of the transitivity clausure adjacency matrix, the dots represent 
the binary value ``1'', the white represent ``0'', it can be notice that the 
complete graph is not dense.}
\label{fig:2}                                    
\end{figure}

 In order to calculate these parameters for real cases, non connected graphs were
 taken into account, that means that is not dense the closure adjacency matrix obtained
from the Warshall algorithm. This can be observed in fig-\ref{fig:2}.
This show different values from the averaged geodesic due to the fact that some actors
are not linked. Therefore we adopted an {\em ad hoc} criterion.
The parameters were calculated in the maximal dense subgraphs where are the
more popular actors which is according with {\em in situ} observations.
That is, the behavior of a mailing list is given by the more related 
actors and not by the isolated or casuals answerers. 
Because of this the number of vertices is reduced 
in 60\% which have no relevance due to the huge number of actors.

\begin{figure}[h]                                                               
\begin{center}                                                                  
\includegraphics[width=12cm, height=13cm,angle=-90]{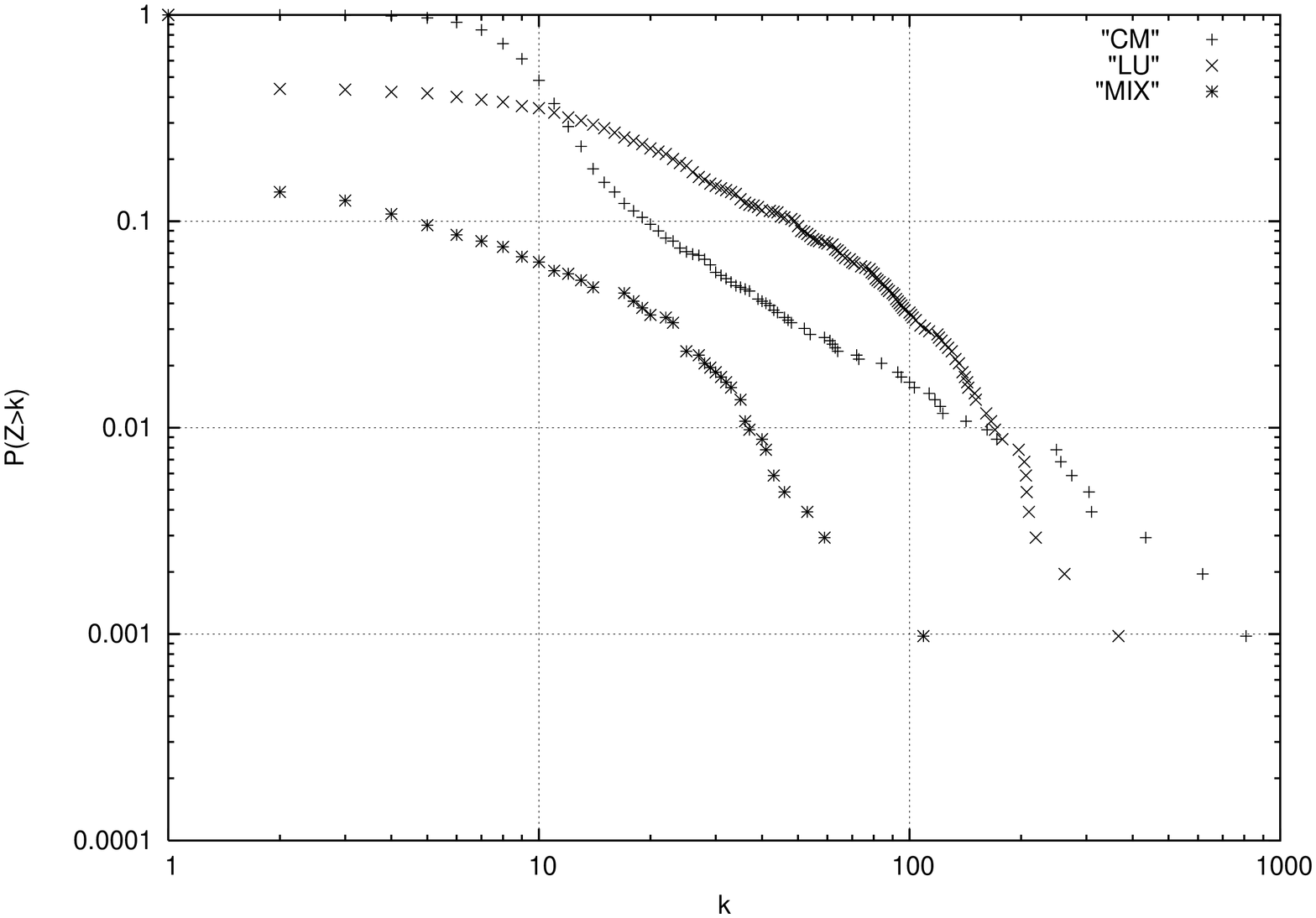}
\end{center}                                                                    
\caption{Log plot of $F(k)$, where in any case $F(0)=1$. On the other hand {\bf LU}
and {\bf MIX} show an abrupt leap due to the abundance of impopular actors.}
\label{fig:3}                                    
\end{figure}

The cumulative probability $F(k)$ is the most significative evidence of the 
difference between the null hypothesis and the real cases. 
As can be seen in fig-\ref{fig:3} 
the tail ($k\to\infty$) is quite different from the theoretical straight line for
{\bf CM} as in``{\em scale free}'' networks. The behavior is ``{\em single-scale}''.

\begin{figure}[H]                                                             
\begin{center}                                                                  
\includegraphics[width=12cm, height=13cm,angle=-90]{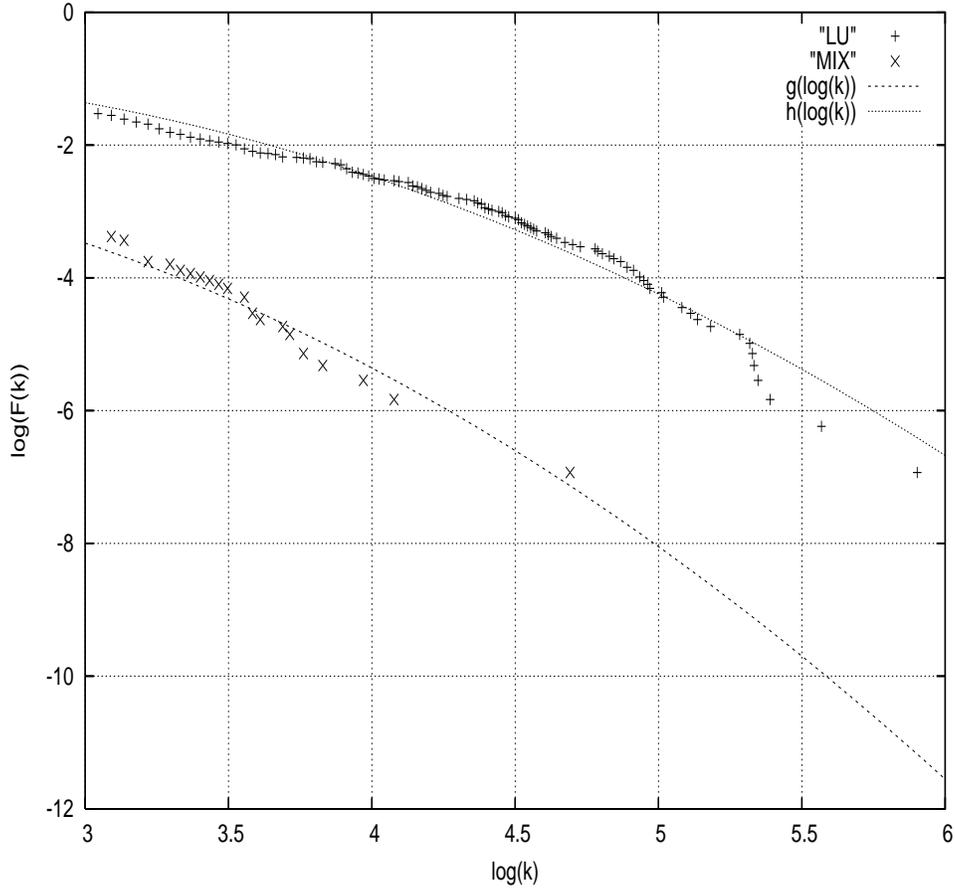}
\end{center}                                                                    
\caption{Logarithmic plot where can be observed the fitting with the data.}
\label{fig:4}                                    
\end{figure}

Instead of perform a linear fitting\footnote{following Amaral {\em et al\@.}\cite{r7}
in an intermediate range an without taking into account the tail.}
We use a quadratic fitting $\log(F(k))=\alpha - \gamma\,(\log(k))^2$. 
In fig-\ref{fig:4} may be notice the goodness of the fitting in the tail 
as the intermediate range. This give us the possibility of discard the behavior
$P(k)\sim k^{-\gamma}$ of the tail and replace for another
$P(k)\sim k^{-\gamma\,\log(k)}$. Where $\gamma=0.17$ for {\bf LU} 
and $\gamma=0.24$ for {\bf MIX}.

In worth to notice the same actors are linked in different way if the themes are 
different, in this case is not the same the tail for the list {\bf LU}
than for the list {\bf MIX}. Therefore the language itself it not the unique constraint
in the network behavior although the social paradigm where the actors 
are involved.
This also indicates that the cumulative probability would be considered as a
qualification element of the social behavior in this kind of societies.

\section{Conclusions}

In this work we concluded that the social relations among a set of identical actors
is strongly linked with the social paradigm where they are involved. 
On the other hand, at least in the societies under analysis the tail
behavior $F(k)$, allows to quantify their differences. 
We may speculate and think that the dissociative character of these societies
may be attributed to the fact these are open societies instead of closed ones 
as ``{\em small-world}. That means that no all the actors 
are related themselves by answering the mails and some actors cause 
the extinction of a theme by avoiding any close link. 

\bibliographystyle{apsrev}

\end{document}